\documentclass[usenatbib]{mn2e}

\usepackage{savesym}
\usepackage{amsmath}
\savesymbol{iint}
\usepackage{txfonts}
\restoresymbol{TXF}{iint}

\usepackage{subfigure}
\usepackage{epsfig}
\usepackage{url}
\def\simless{\mathbin{\lower 3pt\hbox
{$\rlap{\raise 5pt\hbox{$\char'074$}}\mathchar"7218$}}}   %< or of order
\def\simmore{\mathbin{\lower 3pt\hbox
{$\rlap{\raise 5pt\hbox{$\char'076$}}\mathchar"7218$}}}   %> or of order
\newcommand{\be}{\begin{equation}}
\newcommand{\ee}{\end{equation}}
\newcommand       \bea          {\begin{eqnarray}}
\newcommand       \eea          {\end{eqnarray}}
\newcommand       \apj          {ApJ}
\newcommand       \apjl         {ApJL}
\newcommand       \aap          {A\&A}
\newcommand       \nat          {Nature}
\newcommand       \mnras        {MNRAS}

\newcommand       \aj      {AJ}
\newcommand       \prd      {Phys.~Rev.~D.~}

\def\simlt{\mathrel{\hbox{\rlap{\hbox{\lower4pt\hbox{$\sim$}}}\hbox{$<$}}}}
\def\simgt{\mathrel{\hbox{\rlap{\hbox{\lower4pt\hbox{$\sim$}}}\hbox{$>$}}}}
\def\lesssim{\mathrel{\hbox{\rlap{\hbox{\lower4pt\hbox{$\sim$}}}\hbox{$<$}}}}

\def\gtrsim{\mathrel{\hbox{\rlap{\hbox{\lower4pt\hbox{$\sim$}}}\hbox{$>$}}}}

\title[Radio constraints on neutron star merger remnants]{Constraints on long-lived remnants of neutron star binary mergers from late-time radio observations of short duration gamma-ray bursts }

\author[]{Brian~D.~Metzger$^{1}\thanks{E-mail: bmetzger@phys.columbia.edu}$, Geoffrey C.~Bower$^{2}$\\
$^{1}$Department of Physics, Columbia University, New York, NY, 10027, USA\\ $^{2}$Astronomy Department and Radio Astronomy Laboratory, University of California, Berkeley, Berkeley, CA 94720, USA}

\begin{document}
\date{Received / Accepted}
\pagerange{\pageref{firstpage}--\pageref{lastpage}} \pubyear{2013}

\maketitle

\label{firstpage}

\begin{abstract}

The coalescence of a binary neutron star (NS) system (a `NS merger' or NSM) may in some cases produce a massive NS remnant that is long-lived and, potentially, indefinitely stable to gravitational collapse.  Such a remnant has been proposed as an explanation for the late X-ray emission observed following some short duration gamma-ray bursts (GRBs) and as possible electromagnetic counterparts to the gravitational wave chirp.  A stable NS merger remnant necessarily possesses a large rotational energy $\gtrsim 10^{52}$ erg, the majority of which is ultimately deposited into the surrounding circumburst medium (CBM) at mildly relativistic velocities.  We present Very Large Array (VLA) radio observations of 7 short GRBs, some of which possessed temporally extended X-ray emission, on timescales of $\sim 1-3$ years following the initial burst.  No radio sources were detected, with typical upper limits $\sim 0.3$ mJy at $\nu = 1.4$ GHz.  A basic model for the synchrotron emission from the blast wave is used to constrain the presence of a long-lived NSM remnant in each system.  Depending on the GRB, our non-detections translate into upper limits on the CBM density $n \lesssim 3\times 10^{-2}-3$ cm$^{-3}$ required for consistency with the remnant hypothesis.  Our upper limits rule out a long-lived remnant in GRB 050724 and 060505, but cannot rule out such a remnant in other systems due to their lower inferred CMB densities based on afterglow modeling or the lack of such constraints.  

%Additional VLA observations in the near future could place tighter limits on the presence of merger remnants in these systems, %although ruling out such a remnant completely will be challenging without better constraints on the CMB density.    The %population of long-lived NSM remnants will also be constrained by their (non-)detection with upcoming radio transient surveys.  

\end{abstract} 
  

\section{Introduction} 
\label{intro}

The coalescence of binary neutron stars [NSs] (hereafter `neutron star mergers', or NSMs) are promising central engines for powering short-duration gamma-ray bursts (GRBs) (\citealt{Paczynski86}; \citealt{Eichler+89}).  NSMs are also the primary source of gravitational waves for upcoming ground-based interferometric detectors such as Advanced LIGO and Virgo (\citealt{Abadie+10}).

Numerical simulations of the merger process (e.g.~\citealt{ruffert1999}; \citealt{Uryu+00}; \citealt{Rosswog&Liebendorfer03};
\citealt{Oechslin&Janka06}; \citealt{Chawla+10}; \citealt{Rezzolla+10};
\citealt{Hotokezaka+11}) show that the end product of a NSM is typically a hypermassive NS remnant, which is (at least temporarily) stable to gravitational collapse as the result of support by thermal pressure and/or differential rotation (\citealt{Morrison+04}; \citealt{OConnor&Ott11}; \citealt{Paschalidis+12}; \citealt{Lehner+12}).  In the past it has generally been assumed that the neutron star collapses to form a black hole within a relatively short timescale $\lesssim 100$ ms.  The newly created black hole is surrounded by a thick remnant torus of mass $M_{\rm t} \sim 10^{-2} M_{\odot}$, the subsequent accretion of which may power the transient relativistic jet responsible for the GRB (e.g.~\citealt{Narayan+92}; \citealt{Rezzolla+11}).  

One of the biggest uncertainties in predicting the outcome of a NSM results from our incomplete knowledge of the equation of state of high density matter \citep{Hebeler+13}.  The recent discovery of massive $\sim 2M_{\odot}$ neutron stars (\citealt{Demorest+10}; \citealt{Antoniadis+13}) indicates that the EoS is not as soft as predicted by some previous models.  This increases the likelihood that NSMs could not only produce hyper-massive NSs, but also NSs which are {\it indefinitely stable to gravitational collapse} (e.g.~\citealt{Ozel+10}; \citealt{Bucciantini+12}; \citealt{Giacomazzo&Perna13}).  Due to the high angular momentum of the merging binary, such a NS remnant will necessarily be rotating extremely rapidly, with a rotation period $P \sim 1$ ms close to centrifugal break-up.  The NS remnant may also acquire a strong magnetic field $\gtrsim 10^{14}-10^{15}$ G similar to those of Galactic `magnetars', since the initially weak field will be amplified by shear instabilities at the merger interface (\citealt{Price&Rosswog06}; \citealt{Zrake&MacFadyen13}) or by an $\alpha-\Omega$ dynamo (\citealt{Duncan&Thompson92}) in the subsequent neutrino cooling phase.

The possibility that some mergers could leave stable magnetar remnants has recently received heightened attention as a possible explanation for the X-ray activity observed after some short GRBs, such as the `extended emission' observed on timescales of minutes after the GRB (\citealt{Metzger+08}; \citealt{Metzger+11}; \citealt{Bucciantini+12}) and the X-ray `plateaus' observed on longer timescales (\citealt{Rowlinson+13}; \citealt{Gompertz+13}).  Accretion of the remnant torus cannot readily explain this emission, since the torus is not expected to be present at such late times due to powerful outflows that occur on timescales of seconds or less (e.g.~\citealt{Fernandez&Metzger13}).  Prolonged energy injection from a magnetar remnant has also been proposed as a potential electromagnetic counterpart to the gravitational wave signal (\citealt{Zhang13}; \citealt{Gao+13}; \citealt{Yu+13}).  A similar configuration, with an especially long-delayed collapse of the neutron star into a BH, has been suggested as a source of bright cosmological radio bursts (\citealt{Falcke&Rezzolla13}).

A stable neutron star remnant formed by a NSM possesses a substantial reservoir of rotational energy
\be
E_{\rm rot} = \frac{1}{2}I\Omega^{2} \simeq 3\times 10^{52}{\rm erg}\left(\frac{P}{1\,\rm ms}\right)^{-2},
\label{eq:Erot}
\ee
where $I \simeq 1.5\times 10^{45}$ g cm$^{2}$ is the NS moment of inertia and $P = 2\pi/\Omega$ is the rotation rate of the neutron star.  Depending on the dipole magnetic field strength $B_{\rm d}$, this energy will be transferred to the surrounding environment via magnetic spin-down on a characteristic timescale of $\sim$ minutes to hours for magnetar-strength fields $B_{\rm d} \sim 10^{14}-10^{15}$ G, or on longer timescales $\sim$ months$-$years for lower strength fields $B_{\rm d} \sim 10^{12}-10^{13}$ G. 

Only a small fraction of $E_{\rm rot}$ is likely to be radiated: the isotropic radiated energies of short GRBs are typically $\sim 10^{49}-10^{51}$ erg $\ll 10^{52}$ erg (e.g.~\citealt{Nakar07}).  The majority of $E_{\rm rot}$ will instead be transferred into surrounding ejecta and, ultimately, the circumburst medium (CBM).  If the magnetar couples its energy effectively to the ejecta surrounding the merger site (\citealt{Bucciantini+12}), this matter will be accelerated to a mildly-relativistic velocity
\be
\beta_0 \equiv \frac{v}{c} \sim \left(\frac{2E_{\rm rot}}{M_{\rm ej}}\right)^{1/2} \sim 1.0\left(\frac{E_{\rm rot}}{10^{52}{\rm erg}}\right)^{1/2}\left(\frac{M_{\rm ej}}{10^{-2}M_{\odot}}\right)^{-1/2},
\label{eq:vej}
\ee
where $M_{\rm ej} \sim 10^{-2}M_{\odot}$ is the ejecta mass, normalized to a characteristic value (e.g.~\citealt{Hotokezaka+13}).  If the magnetar wind does not couple its energy to the ejecta, the velocity of the outflow will be even faster.  

An inescapable prediction of the remnant NS merger model is thus the ejection of $\sim$ few $10^{52}$ erg of quasi-isotropic energy in at least mildly relativistic matter with $\beta_0 \sim 1$.  This matter will produce bright radio emission once its decelerates via its interaction with the CBM (\citealt{Nakar&Piran11}).  The ejecta transfers most of its energy to the CBM at the characteristic radius where it sweeps up a mass comparable to its own, 
\be
R_{\rm dec} \simeq \left(\frac{3E_{\rm rot}}{4\pi n m_p c^{2}\beta_0^{2}}\right)^{1/3} \approx 1.2\times 10^{18}{\rm cm}\left(\frac{E_{\rm rot}}{10^{52}\rm erg}\right)^{1/3}\left(\frac{n}{\rm cm^{-3}}\right)^{-1/3}\beta_0^{-2/3},
\label{eq:rdec}
\ee
as occurs on a timescale
\be
t_{\rm dec} \sim R_{\rm dec}/c\beta_0 \approx 1.2\,\,{\rm yr}\left(\frac{E_{\rm rot}}{10^{52}\rm erg}\right)^{1/3}\left(\frac{n}{\rm cm^{-3}}\right)^{-1/3}\beta_0^{-5/3},
\label{eq:tdec}
\ee
where $n$ is the density of the surrounding CBM.  The shock wave produced by this interaction will generate synchrotron radiation typically peaking on a timescale $t \sim t_{\rm dec}$, as occurs in similar physical contexts such as GRB afterglows and radio supernovae.  

Radio observations on timescales $t \sim t_{\rm dec}$ provide a sensitive probe of the total energy of astrophysical explosion (`radio calorimetry'; e.g.~\citealt{Frail+00}; \citealt{Berger+03}; \citealt{Soderberg+06}; \citealt{Shivvers&Berger11}).  The large value of $t_{\rm dec}$ (eq.~[\ref{eq:tdec}]) implies that late time radio observations are particularly sensitive to the presence of an additional source of mildly relativistic, quasi-isotropic relativistic ejecta.  Although short GRBs have been monitored for radio emission at early times $t \lesssim $ months, such observations at times $\ll t_{\rm dec}$ are not sensitive to the bulk of the energy released if a remnant NS is indeed present in these systems.   

In this paper we present radio observations of several short GRBs taken on timescales of $\sim$ 1-3 years following the burst.   Our search revealed no detections, which as we will show begins to place interesting constraints on the presence of a long-lived NS remnant in these systems.  This paper is organized as follows.  In $\S\ref{sec:data}$ we present our observations.  In $\S\ref{sec:model}$ we describe a simple model for the expected radio light curve of the neutron star remnant ejecta, which we compare to the radio upper limits.  In $\S\ref{sec:conclusions}$ we summarize our conclusions.

\section{Observations}
\label{sec:data}

\begin{table}
\begin{scriptsize}
\begin{center}
\vspace{0.05 in}\caption{Late-Time Radio Upper Limits on Short GRBs}
\label{table:SGRBs}
\begin{tabular}{ccccccc}
\hline \hline
\multicolumn{1}{c}{GRB} &
\multicolumn{1}{c}{$z$} & 
\multicolumn{1}{c}{T$_{\rm obs}$} & 
\multicolumn{1}{c}{F$_{\rm lim}$$^{(a)}$} & 
\multicolumn{1}{c}{n$^{(b)}$} &
\multicolumn{1}{c}{n$_{\rm max}$ $^{(c)}$} & 
\multicolumn{1}{c}{n$^{*}_{\rm max}$ $^{(d)}$}  \\
\hline
 &  & ({\rm d}) & ({\rm mJy}) & (cm$^{-3}$) & (cm$^{-3}$) & (cm$^{-3}$)\\
\hline 
\\
050709 & 0.16 & 924  & 0.35  & $\sim 10^{-4}-0.1$ & 0.03[0.1] & $1[3]\times 10^{-3}$\\
050724 & 0.258 & 913  & 0.24 & $\sim 0.1-10^{3}$ & 0.05[0.2] & $2[5]\times 10^{-3}$ \\
051221A & 0.5465 & 759 & 0.21 & $0.5-2.4\times 10^{-3}$ & 0.2[0.5] & $5[10]\times 10^{-3}$\\
051227 & 0.80 & 753& 0.24 & - & 0.3[0.9] & $8[20]\times 10^{-3}$ \\
060313 & 0.75 ($<$ 1.1) & 677& 0.51& $\sim 10^{-4}$ & 0.6[2] & 0.01[0.04] \\
060505 & 0.089 & 624 & 0.33 & $\sim 1$ & 0.03[0.08] & $6[20]\times 10^{-4}$\\
070714B & 0.9230 & 189 & 0.19  & - & 4[12] & 0.01[0.04]\\
 \\
  \\
\hline
\hline

\end{tabular}
\end{center}
$^{(a)}$Flux density upper limit based on our radio non-detections, estimated as 3 times the rms noise level at the GRB position.
$^{(b)}$Density of the circumburst medium based on GRB afterglow modeling from the literature.  $^{(c)}$Maximum CBM density in remnant neutron star scenario allowed so as not to exceed radio upper limit based on our VLA observations using our fiducial blastwave model (Fig.~\ref{fig:LCs}; Fig.~\ref{fig:constraints}, top panel), calculated for $\epsilon_B = 0.1[0.01]$. $^{(d)}$Maximum CBM density in remnant neutron star scenario allowed so as not to exceed radio upper limit based on hypothetical observations taken with the upgraded VLA (\citealt{Perley+11}) in the near future (September 2014) using our fiducial blastwave model (Fig.~\ref{fig:LCs}; Fig.~\ref{fig:constraints}, bottom panel), calculated for $\epsilon_B = 0.1[0.01]$.  References for the redshifts and CBM density constraints are given in the text.
\end{scriptsize}
\end{table}

Seven short GRBs (Table \ref{table:SGRBs}) were chosen to be observed based on favorable northerm hemisphere sky positions, with a preference for those showing evidence for temporally extended X-ray emission (\citealt{Norris&Bonnell06}), since one motivation for these observations was to test the hypothesis that these events are produced by a long-lived magnetar remnant \citep{Metzger+08}.  

Observations of the selected GRBs were obtained on 19 and 23 January 2008 with the Very Large Array (VLA).  The radio frequency was 1.425 GHz.  The correlator was configured for wideband continuum observations with a total of 100 MHz bandwidth in each of right and left circular polarization.  The VLA was in B configuration producing synthesized beams with a minimum of 3 arcsec in
one axis and ranged from 5 to 18 arcsec in the second axis, depending on declination and the hour angle of the source.  Targets were observed for approximately 20 minutes bracketed by phase calibrator sources.  3C 48 was used as an absolute amplitude calibration source.  Standard interferometric calibration and imaging techniques were applied with the AIPS package \citep{2003ASSL..285..109G}.  No sources were detected at the positions of the optical afterglows for the GRBs.  In Table~1, we provide a $3\sigma$ upper limit for the flux density of the sources.  Thresholds were higher than expected from thermal statistics due to the presence of bright sources in the primary beam and, in some cases, extended galactic emission.

We now briefly summarize the relevant properties of each GRB.  When available, the observed X-ray, optical, and radio afterglow emission from each burst have in some cases been used to constrain the CBM density, assuming the afterglow is synchrotron radiation produced by the collimated GRB outflow.  In our picture the jet responsible for the GRB is either (1) independent of the merger remnant (e.g.~it is powered by accretion onto the central object immediately following the merger), or (2) is also powered by the spin-down of the remnant, but contains only a small fraction of its total rotational energy, due e.g.~to the difficulty of collimating the magnetar wind into a bipolar jet given the low quantity of mass surrounding the merger site (\citealt{Bucciantini+12}).

\subsection*{GRB 050709}

GRB 050709 was one of the first short GRBs to have its host galaxy identified (\citealt{Fox+05}; \citealt{Hjorth+05}; \citealt{Villasenor+05}), with redshift of $z = 0.16$. \citet{Panaitescu06} modeled the early optical afterglow, constraining the CBM density to be $10^{-4} < n < 0.1$ cm$^{-3}$.  This burst was accompanied by temporally extended X-ray emission (e.g.~\citealt{Barthelmy07}).

\subsection*{GRB 050724}

GRB 050724 had an early-type host galaxy at redshift $z = 0.258$ (\citealt{Berger+05}).  The radio afterglow was detected with a flux density $\sim 0.2-0.5$ mJy over the first $\sim$ day (\citealt{Berger+05}), before decaying to a undetectable level over the next week.   Afterglow modeling by \citet{Panaitescu06} constrained the CBM density to be $0.1 < n < 10^{3}$ cm$^{-3}$.    

\subsection*{GRB 051221A}

GRB 051221A (\citealt{Cummings+05b}) occurred in a star-forming galaxy at redshift $z = 0.5465$ (\citealt{Berger&Soderberg05}).  The GRB radio afterglow was detected at $t \sim 1$ day at $\sim 0.1$ mJy, but then decayed to an undetectable level over the next couple weeks \citep{Soderberg+06}.  Afterglow modeling by \citet{Soderberg+06} constrained the CBM density to the range to be $n \sim 0.5-2.4\times 10^{-3}$ cm$^{-3}$.

\subsection*{GRB 051227}
       
GRB 051227 was a short GRB with extended emission (\citealt{Barthelmy+05a}) and no radio detection on a timescale of a few days with an upper limit $\sim 0.1$ mJy (\citealt{Frail05}).  The redshift of the host galaxy identified by \citet{DAvanzo+09} is uncertain, but \citet{DAvanzo+09} suggest a redshift $z \sim 0.8$ (which we adopt) based on matching the colors of the host to the expected luminosity of a similar galaxy type.  To our knowledge no detailed modeling of the optical afterglow or resulting constraints on the CBM density are available.

\subsection*{GRB 060313}

\citet{Roming+06} obtain a photometric redshift $z = 0.75$ ($z < 1.1$ with 90$\%$ confidence) for GRB 060313, which we adopt. \citet{Roming+06} also model the afterglow, finding a low CBM density $n \sim 10^{-4}$ cm$^{-3}$. No radio emission was detected at $t \sim 2$ days, with an upper limit of $0.11$ mJy (\citealt{Soderberg&Frail06}).

\subsection*{GRB 060505}

GRB 060505 occured in a star forming galaxy at $z = 0.089$ (\citealt{Fynbo+06}; \citealt{Ofek+07}).  Based on the stellar population near its position in the host galaxy (\citealt{Thone+08}) and spectral lag (\citealt{McBreen08}), GRB 060505 has been argued to be a member of the long GRB class (although see \citealt{Levesque&Kewley07}), despite its short duration and lack of an observable supernova.  Here we adopt the standard assumption that GRB 060505 was indeed a short burst.  Afterglow modeling by \citet{Xu+09} indicates an CBM density $n \sim 1$ cm$^{-3}$.

\subsection*{GRB 070714B}

GRB 070714B has one of the highest spectroscopically confirmed redshifts $z = 0.9230$ of any short GRB (\citealt{Graham+09}).  No radio afterglow was detected, with an upper limit $\sim 0.1$ mJy on a timescale of $\sim$ 2 weeks (\citealt{Chandra&Frail07}).  This burst was accompanied by temporally extended X-ray emission (e.g.~\citealt{Ibrahim+08}).

%\begin{figure}
%\includegraphics[width=0.5\textwidth]{flux.eps}
%\epsfig{file=constraints2.ps}
%\caption{Example radio light curves for the synchrotron emission from a NSM remnant blast wave at redshift $z = 0.5$.  Models %are shown for two values of the CBM density, $n = 1$ cm$^{-3}$ ({\it blue}) and $ n = 10^{-2}$ cm$^{-3}$ ({\it red}), and for %two assumptions about the fraction of the shocked thermal energy deposited into the magnetic field, $\epsilon_B = 0.1$ ({\it %solid}) and $\epsilon_B = 0.01$ ({\it dotted}).  Vertical dashed lines show the times of our short GRB observations.   } 
%\label{fig:flux}
%\end{figure}

\section{Constraints on Stable Merger Remnant}

\label{sec:model}

\subsection{Blast Wave Model}

\begin{figure*}
\includegraphics[width=0.8\textwidth]{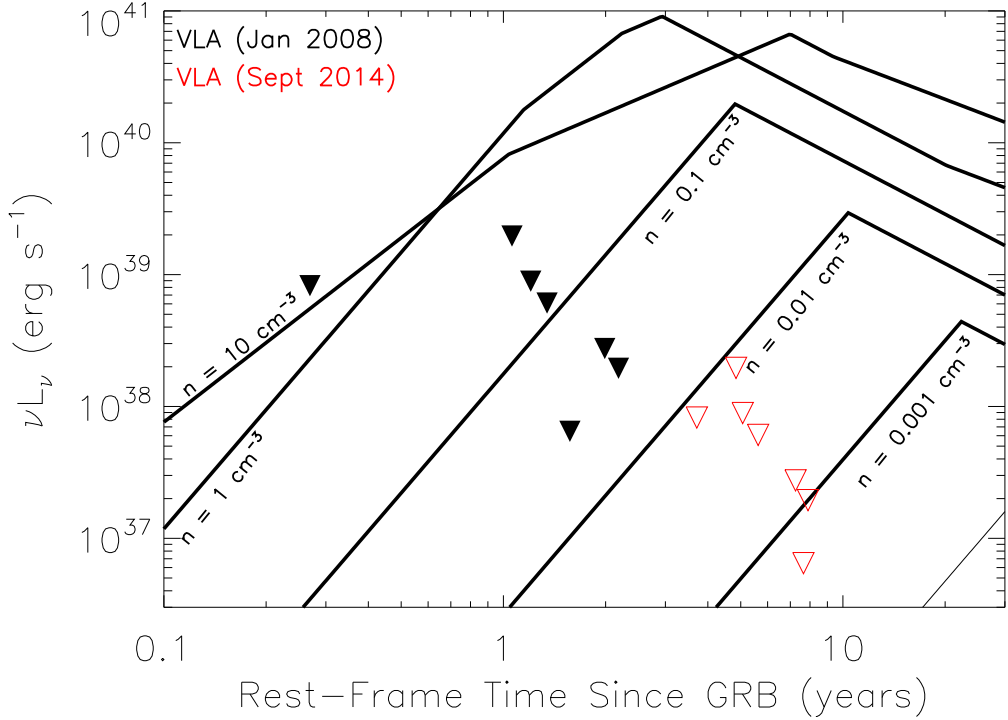}
\caption{Luminosity of synchrotron emission $\nu L_{\nu}$ at $\nu = 1.4$ GHz due to the blast wave produced by the remnant NS as a function of time (rest frame) after the merger, shown for several values of the CBM density $n$ and calculated assuming $\epsilon_e = \epsilon_B = 0.1$.  Shown with black solid triangles are the upper limits from our VLA observations of short GRBs.  Shown in red open triangles are the sensitivity limits that could be obtained by a hypothetical observation in the near future (September 2014) with the upgraded Karl. J. Jansky VLA at $\nu = 1.4$ GHz, assuming a factor $10$ times greater flux density sensitivity than our original VLA observations.} 
\label{fig:LCs}
\end{figure*}

We model the late-time radio emission due to energy injection by a long-lived stable remnant using the synchrotron blast wave model of \citet{Nakar&Piran11}.  Accounting for the early free-coasting phase, and the subsequent Sedov-Taylor expansion, the blast wave velocity $v = \beta c$ evolves as
\begin{eqnarray}
\beta = \left\{
\begin{array}{lr}
\beta_0
, 
\,\,\,\,\,\,\,\,\,\,\,\,\,\,\,\,\,\,\,\,\,\,\,\,\,\,\,\,\,\,\,\,\,\,\,\,\,\,&(r< R_{\rm dec} ) \\
\beta_0(r/R_{\rm dec})^{-3/2},       
\,\,\,\,\,\,\,\,\,\,\,\,\,\,\,\,\,\,\,\,\,\,\,\,\,\,\,\,&(r \ge R_{\rm dec}), \\
\end{array}
\right.
\label{eq:tbo2}
\end{eqnarray}
where $\beta_0$ is the initial velocity and $R_{\rm dec}$ is the deceleration radius (eq.~[\ref{eq:rdec}]).  We adopt a value of $\beta_0 = 0.8$ corresponding to mildly relativistic ejecta (bulk Lorentz factor $\Gamma_0 = (1-\beta_0^{2})^{-1/2} \approx 1.5$).  This value is motivated by our expectation that the majority of the energy is coupled to the relatively massive ejecta (eq.~[\ref{eq:vej}]), resulting in a mildly relativistic quasi-spherical outflow.  Note that since the mass ejected by a NSM is unlikely to exceed $\sim$ few $10^{-2}M_{\odot}$ (e.g.~\citealt{Hotokezaka+13}), equation (\ref{eq:vej}) requires that the ejecta be at least mildly relativistic, $\beta_0 \sim 1$ for $E_{\rm rot} = 3\times 10^{52}$ erg.  If, on the other hand, a large fraction of $E_{\rm rot}$ were instead channeled into a collimated ultra-relativistic jet with $\Gamma_0 \gg 1$, this would produce extremely bright breamed radio emission peaking at earlier times, which is incompatible with the modest radio flux densities of short GRBs measured on timescales of $\sim$ days-weeks after the burst ($\S\ref{sec:data}$) 

%However, even if a portion of the pulsar wind couples to a collimated ultra-relativistic jet, the jet material will slow to become %mildly relativistic and quasi-isotropic on timescales comparable to the deceleration timescale $t_{\rm dec}$ (eq.~[\ref{eq:tdec}]; %\citealt{Zhang&MacFadyen09}), thus producing similar radio emission to our model assuming $\beta_0 = 0.8 $.\footnote{This %behavior is the source of the late radio GRB `orphan' afterglows (\citealt{Rhoads97}; \citealt{Levinson+02}).}    

Input to the radio model include the fraction of the blast wave energy going into relativistic electrons $\epsilon_e$ and the magnetic field $\epsilon_B$, respectively.  We adopt a value of $\epsilon_e = 0.1$, as motivated by the similar values found by modeling GRB afterglows (\citealt{Panaitescu&Kumar02}) and mildly relativistic blast waves from jetted tidal disruption events (\citealt{Metzger+12}; \citealt{Berger+12}).  We adopt a fiducial value of $\epsilon_B = 0.1$, although we explore the sensitivity of our upper limits on the density to the value of $\epsilon_B$ by also considering a lower value $\epsilon_B = 0.01$ in some of our models.  The power-law index of the relativistic electrons is assumed to be $p = 2.3$.  We assume a blast wave energy equal to the rotational energy of the pulsar $E = E_{\rm rot} = 3\times 10^{52}$ erg, since a rotation period near centrifugal break-up appears to be a robust property of stable merger remnants (\citealt{Giacomazzo&Perna13}).  We include corrections to the late radio emission based on the recent work by \citet{Sironi&Giannios13}, although in practice this has little effect on our conclusions.  Our model also includes the effects of synchrotron self-absorption, as described in \citet{Nakar&Piran11}.

%Figure \ref{fig:flux} shows the predicted $\nu = 1.4$ GHz light curve, for different values of $\epsilon_B = 0.01, 0.1$ and $n = %10^{-3}, 0.1$ cm$^{-3}$.

If the observing frequency is above both the peak synchrotron frequency and the self-absorption, as is usually satisfied at $\nu = 1.4$ GHz, then the peak flux density achieved at $t \sim t_{\rm dec}$ is given by (\citealt{Nakar&Piran11})
\be
F_{\rm peak} \approx 3{\rm mJy}\left(\frac{E_{\rm rot}}{10^{52}\rm erg}\right)\left(\frac{n}{\,{\rm cm^{-3}}}\right)^{0.83}\left(\frac{\epsilon_B}{0.1}\right)^{0.83}\left(\frac{\epsilon_e}{0.1}\right)^{1.3}\beta_0^{2.3}d_{28}^{-2},
\label{eq:Fpeak}
\ee
where $d = d_{28}10^{28}$ cm is the luminosity distance, normalized to a value characteristic of the typical redshift $z \sim 0.5$ of short GRBs.  Thus, for typical parameters $E_{\rm rot} = 3\times 10^{52}$ erg, $\epsilon_e = \epsilon_B = 0.1$, $\beta = 0.8$, $d_{28} = 1$, a stable remnant NS should be detectable by the old VLA[upgraded VLA] at $t \sim t_{\rm dec}$ for a CBM density $n \gtrsim 0.02[0.001]$ cm$^{-3}$, where we have assumed a flux density sensitivity of 0.2[0.02] mJy for the old and upgraded VLA, respectively.

%\begin{figure}
%\includegraphics[width=0.8\textwidth]{constraints.eps}

%\caption{Predicted 1.4 GHz radio flux from the blast wave produced by a long-lived NS remnant divided by our radio limits, %calculated as a function of the unknown CBM density $n$ for each GRB as marked, assuming two values of the $\epsilon_B = %0.1$ ({\it solid}) and $\epsilon_B = 0.01$ ({\it dashed}).  If the actual CBM density exceeds the value above which $F \gtrsim %F_{\rm lim}$ ({\it horizontal dashed line}) then a stable remnant can be excluded for that system.  Shown with horizontal lines are %the CBM densities inferred based on independent afterglow modeling for GRB 050709 \citep{Panaitescu06}, GRB 050724 %\citep{Panaitescu06}, GRB 051221A \citep{Soderberg+06}, and GRB 060313 \citep{Roming+06}. } 
%\label{fig:constraints}
%\end{figure}

\subsection{Comparison to Upper Limits}

Figure \ref{fig:LCs} shows the luminosity of synchrotron emission $\nu L_{\nu}$ at frequency $\nu = 1.4$ GHz as a function of time (rest frame) after the merger, shown for several values of the CBM density $n$ and calculated assuming $\epsilon_e = \epsilon_B = 0.1$.  Black filled triangles show the upper limits from our VLA observations of short GRBs.  In most cases the observing band is above the characteristic sychrotron peak frequency, such that the light curve rises to a peak flux density $\sim F_{\rm peak}$ (eq.~[\ref{eq:Fpeak}]) on a timescale $\sim t_{\rm dec}$ (eq.~[\ref{eq:tdec}]).  The different light curve shape in the high density cases $n = 1, 10$ cm$^{-3}$ arises due to the effects of synchrotron self absorption, which is important because the self-absorption frequency is comparable to the observing frequency, especially at early times.   

\begin{figure}
\subfigure{
\includegraphics[width=0.5\textwidth]{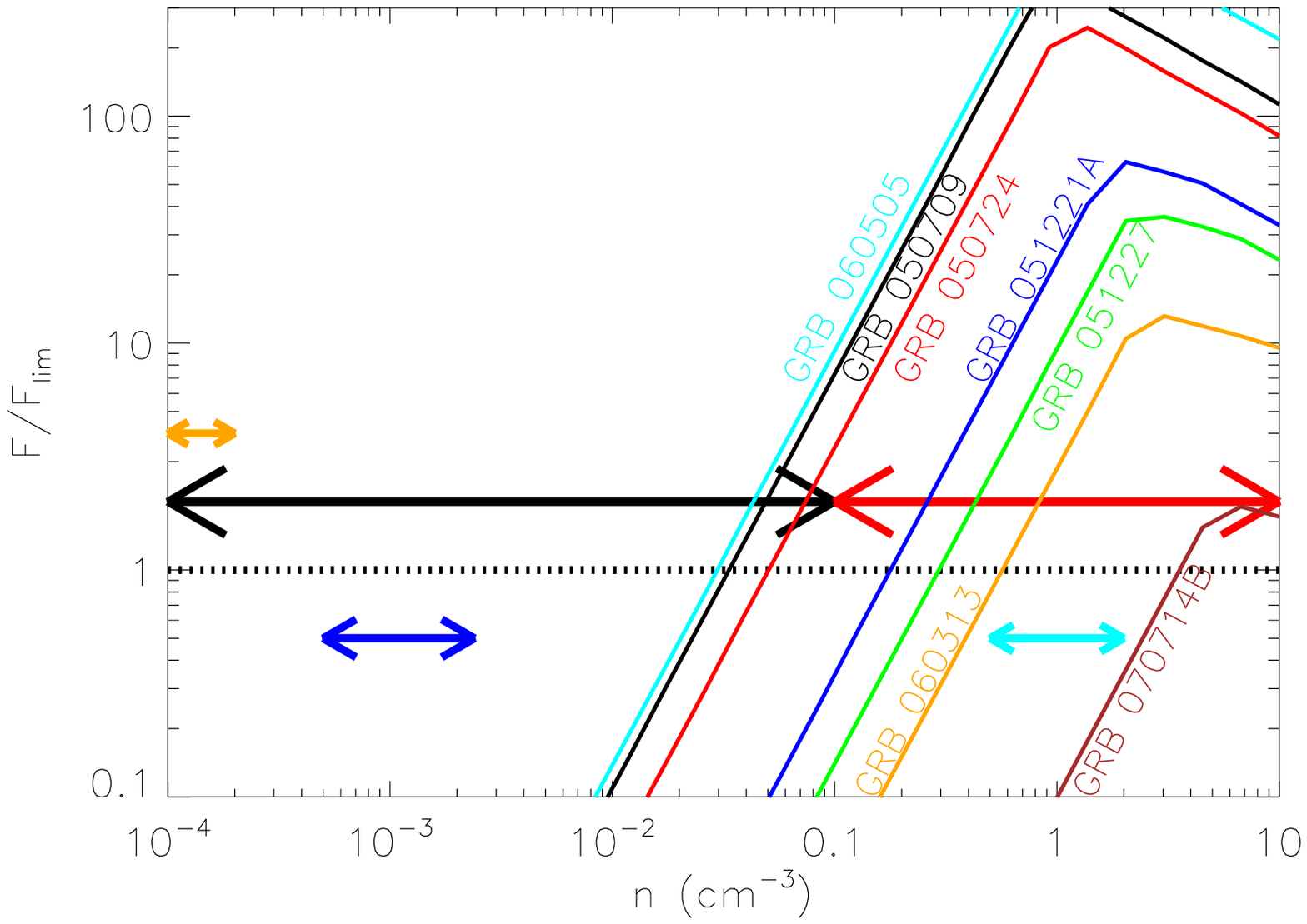}}
\subfigure{
\includegraphics[width=0.5\textwidth]{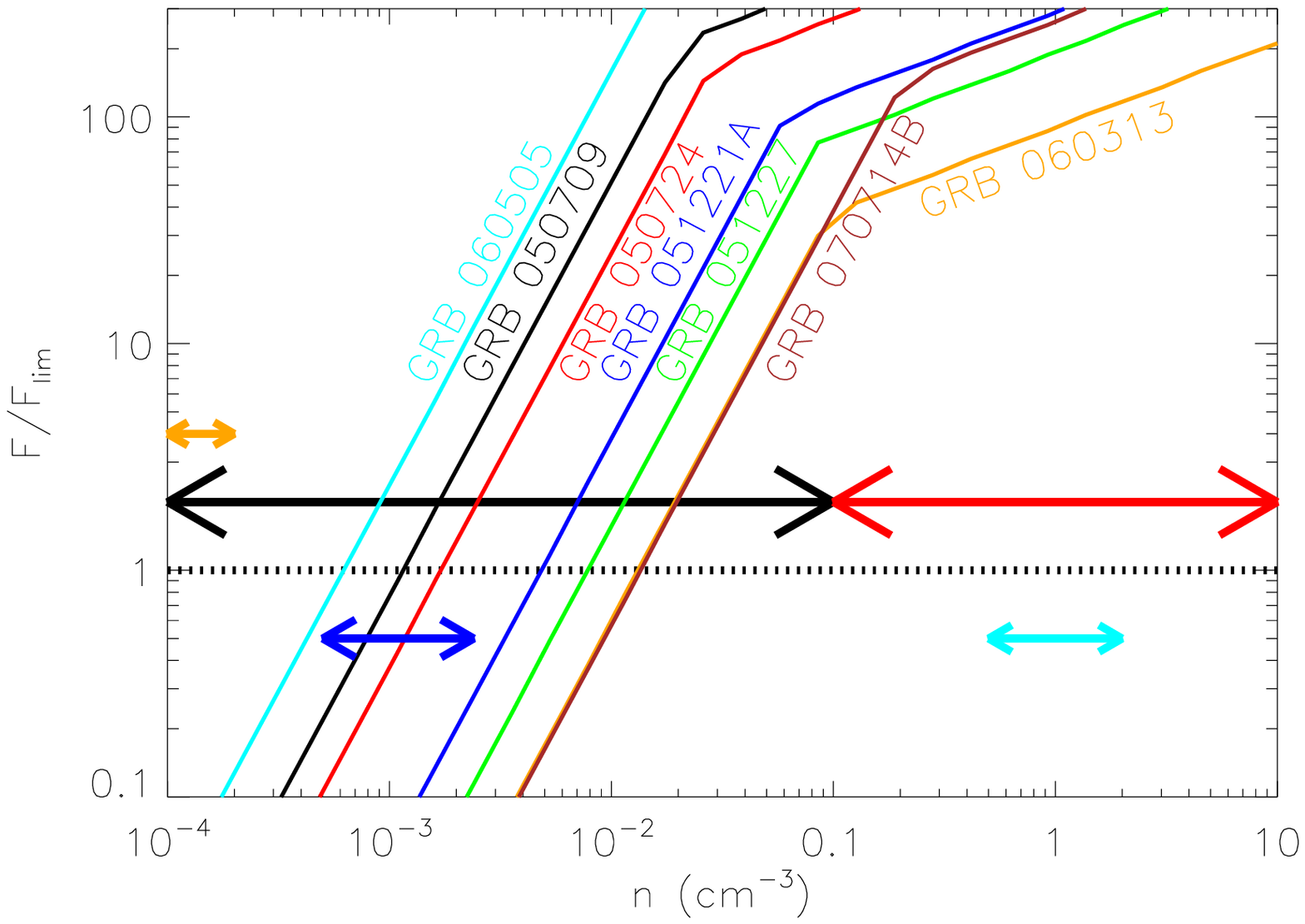}}
\caption{{\large Top Panel:} Predicted 1.4 GHz radio flux density from the blast wave produced by a long-lived NS remnant divided by our radio limits, calculated as a function of the unknown CBM density $n$ for each GRB as marked assuming $\epsilon_e = \epsilon_B = 0.1$ (constraints for the $\epsilon_B = 0.01$ case are given in Table \ref{table:SGRBs}).  If the actual CBM density exceeds the value above which $F \gtrsim F_{\rm lim}$ ({\it horizontal dashed line}) then a stable remnant can be excluded for that system.  Shown with horizontal bars are the CBM densities inferred based on independent afterglow modeling for GRB 050709 \citep{Panaitescu06}, GRB 050724 \citep{Panaitescu06}, GRB 051221A \citep{Soderberg+06}, GRB 060313 \citep{Roming+06}, and GRB 060505 \citep{Xu+09}. {\large Bottom Panel:}
Same as top panel, but calculated based on hypothetical observations in the near future (September 2014) with the VLA at 1.4 GHz.  The upgraded VLA is assumed to achieve a flux density sensitivity which is a factor of 10 lower than for our (original) VLA observations.} 
%For LOFAR we have assumed a $5\sigma$ flux limit of $F_{\nu} = 0.2$ mJy for a 12 hour integration. 
\label{fig:constraints}
\end{figure}

Figure \ref{fig:LCs} shows that we can rule out a long-lived remnant for those short GRBs that occur in a reasonably high density environment $n \gtrsim 0.01-1$ cm$^{-3}$ characteristic of the ISM of star-forming galaxies.  Figure \ref{fig:constraints} provides a different perspective on the data by showing the ratio of the predicted radio flux density of the blast wave at the time of our observations to our radio upper limits for $\epsilon_B = 0.1$.  Horizontal bars show the range of allowed CBM densities for each burst (colored accordingly), based on modeling of the GRB afterglow from the literature (no afterglow modeling was available for GRB 051227 or 070714B).  The maximum CBM density allowed for each GRB so as not to exceed radio upper limit based on our VLA observations is given in Table \ref{table:SGRBs} for both $\epsilon_B = 0.01$ and $\epsilon_B = 0.1$.

Our upper limits rule out a magnetized remnant NS in GRB 050724 and 060505.  However, the low values of $n$ for the other GRBs or the lack of such information, preclude similarly definitive constraints.  Figure \ref{fig:LCs} makes clear that a NSM remnant cannot be ruled out at lower densities due mostly to the fact that our observations were taken at times $t \lesssim t_{\rm dec}$ (eq.~[\ref{eq:tdec}]), when the radio light curve is still rising.  This implies that later observations will be more constraining.  The red triangles in Figure \ref{fig:LCs} show what constraints could be placed by a hypothetical observation in the near future (September 2014) with an instrument such as the upgraded VLA, with a sensitivity assumed to be 10 times greater than the (original) VLA observations.  These same constraints are shown in the bottom panel of Figure \ref{fig:constraints} and in the last column of Table \ref{table:SGRBs}.

\section{Discussion and Conclusions}
\label{sec:conclusions}

A long-lived magnetized NSM remnant has received recent attention as a possible source of X-ray emission following short GRBs and as possible gravitational wave counterparts.   Any NS remnant that indeed avoids collapse is necessarily born rapidly rotating due to the substantial angular momentum of the initial binary.  Although a minimum rotation period of $P \sim 1$ ms may be enforced by gravitational radiation due to non-axisymmetric (e.g.~bar mode) instabilities (e.g.~\citealt{Ott+05}), this still places a robust minimum on the reservoir of rotational energy of $E_{\rm rot} \gtrsim 3\times 10^{52}$ erg  (eq.~[\ref{eq:Erot}]).  This enormous energy must eventually be injected into the surrounding CBM.  Unless the magnetic dipole field of the remnant neutron star is relatively low $\lesssim 3\times 10^{12}$ G, its spin-down time will be shorter than the time required for the blast wave to decelerate via its interaction with the CBM $t_{\rm dec}$ (eq.~[\ref{eq:tdec}]).  The full value of $E_{\rm rot}$ must thus be used as the energy of the blast wave to compute the expected radio emission. 

Motivated by the potential for such lumious radio emission, we undertook VLA observations of several short GRBs, three of which possessed temporally extended X-ray emission.  Although we find no detections, our upper limits are nevertheless constraining on the presence of NSM remnants in these systems.  We can already rule out long-lived NSM remnants in a few short GRBs (050724, 060505) with known high CBM densities, and are placing upper limits on the allowed range of CBM densities in other systems consistent with the NSM remnant hypothesis (Figs.~\ref{fig:LCs}, \ref{fig:constraints}).  One of these bursts, GRB 050724, showed temporally extended X-ray emission; thus our observations are already in tension with models invoking the presence of millisecond magnetars in these systems (\citealt{Metzger+08}; \citealt{Bucciantini+12}; \citealt{Gompertz+13}).

Our upper limits are unconstraining for low CBM densities not because the peak luminosity $F_{\rm peak}$ (eq.~[\ref{eq:Fpeak}]) is too low, but instead because the observations were taken before the light curve has peaked at $t_{\rm peak} \sim t_{\rm dec} \propto n^{-1/3}$ (Fig.~\ref{fig:LCs}).  Since our observations were taken over 5 years ago, additional data taken in the present epoch with more sensitive instruments such the upgraded VLA or LOFAR could substantially tighten the constraints.  The bottom panel of Figure \ref{fig:constraints} shows the constraints on the predicted flux density as a function of CBM density that could be placed by upgraded VLA ($\nu = 1.4$ GHz) by observations taken in September 2014.  Unfortunately, given the low or uncertain CBM densities in the remaining GRBs in our sample, it does not appear that additional non-detections would allow us to rule out a remnant merger in events beyond those already constrained (GRB 050724 and 060505).  We nevertheless expect that additional, more recent, short GRBs could be added to our sample for future observations, prioritized by systems with known high densities.  

Although particularly low CBM densities hinder our ability to constrain NSM remnants in individual systems, such low densities should not be ubiquitous among the population of binary neutron star mergers.  Although some NSMs are predicted to occur outside of their host galaxies due to the natal kicks received by the NS at birth (e.g.~\citealt{Belczynski+06}; \citealt{Kelley+10}), this is unlikely to always be the case.  In fact, the low mass NSs most likely to leave stable remnants are produced in electron capture supernovae, for which the expected kicks are much smaller (e.g.~\citealt{Buras+06}).  {\it Thus, most mergers leaving long-lived remnants should occur within the confines of their host galaxies, where the CBM density is relatively high.}   

The constraints our upper limits place on the presence of temporarily stable (`hyper-massive') neutron star remnants are much less strict than those on indefinitely stable remnants.  This is because a hyper-massive neutron star is unlikely to live longer than a few seconds, as set by the time required for thermal support to be removed via neutrino cooling and for differential rotation to be eliminated by magnetic torques.  Only if the dipole magnetic field of the remnant is extremely high $\gtrsim 10^{16}$ G could a substantial fraction of its rotational energy be extracted prior to collapse.  Separately, another caveat is that our radio upper limits are also only constraining on stable remnants that possess a moderately strong dipole magnetic field $B_{\rm d} \gtrsim$ few 10$^{12}$ G, since otherwise the spin-down timescale is too long to power a substantial blast wave luminosity.

It is uncertain theoretically whether a system leaving a NSM remnant can produce a GRB at all due to the substantial baryon loading of the jet expected from the neutrino-driven wind of the NSM remnant (e.g.~\citealt{Dessart+09}).  However, independent of their association with GRBs, the formation of  long-lived remnants in NS mergers is also constrained based on the current absence of such radio transients in blind radio surveys.  

\citet{Frail+12} present updated limits on the rates of transients based on past radio surveys.  They predict that that number of NSM transients per square degrees is approximate two orders of magnitude below the limits of current surveys (\citealt{Frail+03}; \citealt{deVries+04}; \citealt{Croft+10}).  However, the blast wave energy they assume, $E = 10^{50}$ erg (\citealt{Nakar&Piran11}), is more than two orders of magnitude lower than that produced by a long-lived NSM remnant.  The peak flux density of the resulting radio transient will thus be approximately two orders of magnitude brighter (eq.~[\ref{eq:Fpeak}]), increasing the expected rate of detected events at fixed flux density by a factor $\sim 10^{3}$.  

Using Figure 6 of \citet{Frail+12} we estimate that one can exclude the possibility that a majority of NSMs leave stable remnants based on the upper limits placed by past radio surveys, if one assumes a rate of binary neutron star mergers $\sim 300$ Gpc$^{-3}$ yr$^{-1}$ near the middle of the estimated range (\citealt{Abadie+10b}).  This illustrates the power of upcoming radio surveys, combined with future gravitational wave observations, to constrain the fraction of mergers leaving stable remnants and hence indirectly constrain the equation of state of nuclear density matter.  On a related note, the much higher radio luminosities produced in the case of a remnant NS, as compared to the normal case of prompt black hole formation, make radio observations a much more promising electromagnetic counterpart to the gravitational wave source than previously assumed (\citealt{Nakar&Piran11}; \citealt{Metzger&Berger12}).

As a final remark, note that our upper limits, and also those placed by radio surveys, place similarly strong constraints on the proposed long-lived electromagnetic spin-down of {\it black holes} produced by NSMs \citep{Lyutikov13}.  This suggests that such newly-formed black holes indeed rapidly shed their `hair' (\citealt{Lyutikov&McKinney11}).

%\begin{figure}
%\includegraphics[width=0.5\textwidth]{constraints_future.eps}
%\caption{Same as Figure \ref{fig:constraints}, but calculated based on hypothetical observations at the present epoch (September %2013) with LOFAR at 200 MHz ({\it solid}) and upgraded VLA at 1.4 GHz ({\it dashed}).  For LOFAR we have assumed a $5\sigma$ flux %limit of $F_{\nu} = 0.2$ mJy for a 12 hour integration.  For EVLA we have assumed} 
%\label{fig:futureconstraints}
%\end{figure}

%\begin{figure}
%\includegraphics[width=0.5\textwidth]{constraints2.eps}
%\epsfig{file=constraints2.ps}
%\caption{constraints.} 
%\label{fig:constraints}
%\end{figure}

\section*{Acknowledgements}

The National Radio Astronomy Observatory is a facility of the National Science Foundation operated under cooperative agreement by Associated Universities, Inc.  BDM acknowledges support from the Department of Physics at Columbia University.

%\bibliographystyle{mn2e}
%\bibliography{../biblio/bibliography}
%\bibliography{ms}

%\begin{thebibliography}{}
%\end{thebibliography}

\end{document}